\documentclass[aps,pra,amsmath,nofootinbib,twocolumn,longbibliography,floatfix]{revtex4-2}

\RequirePackage{graphicx}
\RequirePackage{hyperref} 
\RequirePackage{isotope}
\RequirePackage{float}
\RequirePackage{ifthen}
\RequirePackage[utf8]{inputenc}
\RequirePackage{newunicodechar} 
\RequirePackage{xcolor}
\RequirePackage{gensymb}
\RequirePackage[T1]{fontenc}

\newunicodechar{ }{ }                       
\newunicodechar{−}{\ensuremath{-}}          
\newunicodechar{∕}{\ensuremath{/}}          
\newunicodechar{×}{\ensuremath{\times}}     
\newunicodechar{°}{\degree}                 

\newcommand{\expn}[1]{\ensuremath{\times 10^{#1}}}

\newcommand{\reffig}[1]{Fig.~\ref{#1}}

\newcommand{\refapp}[1]{Appendix~\ref{#1}}

\newcommand{\ket}[1]{\ensuremath{\left|#1\right\rangle}}

\newcommand{\Si}{\ensuremath{^{29}\text{Si}}}
\newcommand{\Ge}{\ensuremath{^{73}\text{Ge}}}

\newcommand{\nosc}{\ensuremath{N_\text{osc}}}
\newcommand{\tidle}{\ensuremath{t_\text{idle}}}
\newcommand{\tpulse}{\ensuremath{t_\text{pulse}}}

\newcommand{\ttwostar}{\ensuremath{T_{2}^*}}

\newcommand{\cnotF}{$96.3 \pm 0.7\%$}
\newcommand{\swapF}{$99.3 \pm 0.5\%$}
\newcommand{\lcczF}{$93.8 \pm 0.7\%$}
\newcommand{\twoRBF}{$97.1 \pm 0.2\%$}

\begin{document}

\title{Universal logic with encoded spin qubits in silicon}
\date{\today}
\author{Aaron~J.~Weinstein} \email[]{ajweinstein@hrl.com} 
\author{Matthew~D.~Reed}
\author{Aaron~M.~Jones}
\author{Reed~W.~Andrews}
\author{David~Barnes}
\author{Jacob~Z.~Blumoff}
\author{Larken~E.~Euliss}
\author{Kevin~Eng}
\author{Bryan~Fong}
\author{Sieu~D.~Ha}
\author{Daniel~R.~Hulbert}
\author{Clayton~Jackson}
\author{Michael~Jura}
\author{Tyler~E.~Keating}
\author{Joseph~Kerckhoff}
\author{Andrey~A.~Kiselev}
\author{Justine~Matten}
\author{Golam~Sabbir}
\author{Aaron~Smith}
\author{Jeffrey~Wright}
\author{Matthew~T.~Rakher}
\author{Thaddeus~D.~Ladd}
\author{Matthew~G.~Borselli}

\affiliation{HRL Laboratories, LLC, 3011 Malibu Canyon Road, Malibu, California 90265, USA}

\begin{abstract}
Qubits encoded in a decoherence-free subsystem and realized in exchange-coupled silicon quantum dots are promising candidates for fault-tolerant quantum computing.
Benefits of this approach include excellent coherence, low control crosstalk, and configurable insensitivity to certain error sources.
Key difficulties are that encoded entangling gates require a large number of control pulses and high-yielding quantum dot arrays.
Here we show a device made using the single-layer etch-defined gate electrode architecture that achieves both the required functional yield needed for full control and the coherence necessary for thousands of calibrated exchange pulses to be applied.
We measure an average two-qubit Clifford fidelity of \twoRBF\ with randomized benchmarking.
We also use interleaved randomized benchmarking to demonstrate the controlled-NOT gate with \cnotF\ fidelity, SWAP with \swapF\ fidelity, and a specialized entangling gate that limits spreading of leakage with \lcczF\ fidelity.
\end{abstract}

\maketitle

Quantum computers promise hardware acceleration for certain problems but are challenging to build due to the need to satisfy a conflicting set of demands.
Qubits need to be well-isolated from microscopic sources of noise but simultaneously controlled with exquisite analog precision and high speed, all in a platform capable of scaling to sizes of computational relevance.
Critical to these goals is fault tolerance (FT), where information is encoded in a way that contains and negates errors with a combination of redundancy, symmetry, and careful scheduling of operations.
FT is chiefly discussed in the context of active quantum error correction, but given the strict requirements of error rates and correlation required for FT, other error mitigation schemes will certainly be necessary. 
Decoherence-free subsystems (DFS) constitute one likely ingredient since they enable the engineering of desirable qubit properties~\cite{andrews2019} including error insensitivity and reduced temporal or spatial correlation~\cite{bacon2000} by encoding information in multiple imperfect quantum systems.  
Here, we demonstrate universal exchange-only control of two encoded DFS qubits in an array of silicon quantum dots using error-mitigating composite sequences.
These sequences include encoded two-qubit entangling gates, SWAP operations that are designed to move data with reduced sensitivity to our dominant error source of nuclear magnetic noise, and leakage-controlled entangling gates that restrict certain types of errors from spreading.
Together, these results constitute universal two-qubit ``exchange only'' (EO) control well-suited to demonstrations requiring larger qubit arrays.

Significant progress with silicon quantum dot qubits has recently been made~\cite{burkard2021}, in part due to an increased appreciation for their advantages.
Isotopically-enhanced silicon is known to host exquisite spin memory times~\cite{tyryshkin2012, saeedi2013}, indicating the potential for environmental isolation.
Gated structures lithographically defined using semiconductor manufacturing processes can be engineered to deliver control signals precisely.
Fabrication of larger arrays should also follow yield-enhancement trajectories established by the silicon-based classical microelectronics industry as designs are improved~\cite{zwerver2021, kuemmeth2020}. 
Demonstrations with silicon qubits include single-spin control via electron-dipole-spin resonance~\cite{Yoneda2018}, initialization and read-out of entangled spin states~\cite{blumoff2022}, entangling operations between two single-spin qubits using exchange~\cite{xue2022, noiri2022, mills2021}, and charge shuttling in arrays as large as nine dots~\cite{mills2019, yoneda2021}.
However, many of these results have employed control methods which may present challenges for scaling.
Those include microwave control signals which risk cross-talk and phase-tracking issues, microscale ferromagnets generating magnetic field gradients for which scalable array designs present challenges, and the spin-orbit effect of holes which are strongly subject to microscopic device disorder and charge noise~\cite{burkard2021}.

\begin{figure*}[ht!]
	\centering
	\includegraphics[width=\textwidth]{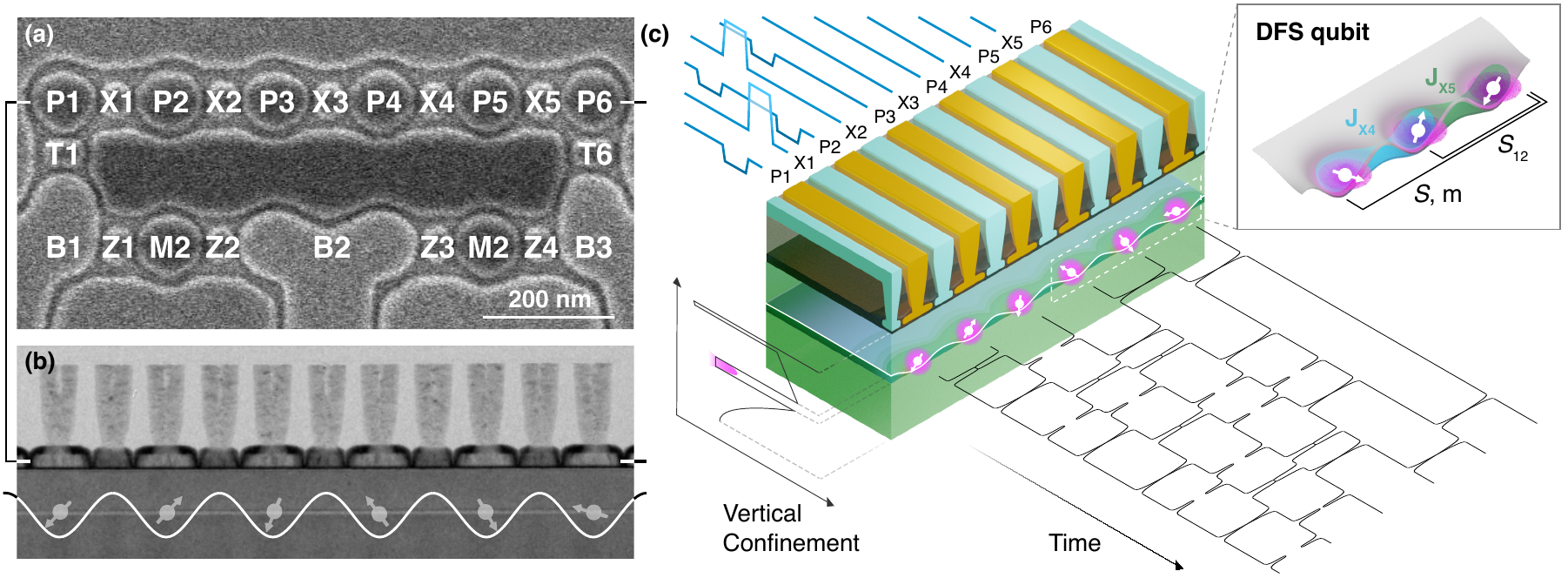}
	\caption{
		A six-quantum-dot, two-qubit SLEDGE device in Si. 
		(a) Top-view SEM micrograph of the metallic gates that comprise the device’s plunger (P), exchange (X), tunnel (T), bath (B), barrier (Z), and measure-dot (M) gates.  
		Two DFS qubits are formed with the P1-P3 and P4-P6 dots and connected by a single exchange gate (X3).
		(b) Cross-sectional TEM micrograph of gate and via electrodes, cut along the dashed line in (a). 
		Electrons are vertically confined by the Si/SiGe heterostructure boundaries and laterally confined by the induced electrostatic potential of the device gates. 
		(c) Qubit states are manipulated with sequences of nearest-neighbor exchange interactions that are principally modulated by X gates \cite{reed2016,andrews2019}.   
		We perform entangling operations in the two-qubit computational subspace with sequences like the one shown in the lower right.
	}
	\label{fig:device}
\end{figure*}

Exclusive use of the exchange interaction is a compelling alternative for spin qubit control~\cite{divincenzo2000}. 
Modulating the gated exchange interaction requires only baseband voltage pulses which can be applied asynchronously, obviating the need for less scalable control mechanisms or precise tracking of rotating frames.  
Exchange is also exponentially suppressed spatially, natively providing low control crosstalk and inherently limiting the error correlations which might otherwise spoil FT.
The cost of this EO control modality is that qubits must be encoded in a minimum of three physical spins and encoded gate operations require multiple discrete exchange pulses~\cite{andrews2019}.
EO two-qubit entangling gates are particularly complicated since they must preserve both qubit encodings without leakage while applying the desired operation via a control trajectory through a much larger state space.
Here the encoded CNOT gate takes 28 pulses with the Fong-Wandzura (FW) construction \cite{fong2011} (see \refapp{app:theory}), compared to only a single exchange pulses for single-spin qubits~\cite{xue2022, noiri2022, mills2021}. 

Though EO control was proposed over twenty years ago~\cite{divincenzo2000}, experimental demonstration of encoded entangling operations has waited for the availability of fully-functional six-dot devices.  
Three recent advances enabled the result we report on here: the Single Layer Etch-Defined Gate Electrode (SLEDGE) process~\cite{haha2021} for making high-yielding devices, the use of narrow Si/Si$_{0.3}$Ge$_{0.7}$ quantum wells to increase the probability that dots have sufficiently large valley energy for measurement and initialization~\cite{blumoff2022}, and improved control software to manage the complexity of these larger quantum manipulations~\cite{andrews2019}.
Using these, we demonstrate an average 2-qubit Clifford fidelity of \twoRBF\ and the FW controlled-NOT gate~\cite{fong2011, zeuch2016} (FW-CNOT) with \cnotF\ fidelity, both limited by well-understood sources of magnetic noise~\cite{kerckhoff2021}, and an encoded SWAP operation with \swapF\ fidelity.
We also demonstrate a ``leakage controlled'' controlled-Z (LCCZ) gate, requiring 45 exchange pulses, with \lcczF\ fidelity.
The LCCZ improves on FW gates by preventing occupation of unencoded ``leaked'' spin states from spreading, a feature which may prove critical to FT.

Our SLEDGE-based device~\cite{haha2021} is composed of six quantum dots arranged in a line, corresponding to two three-spin EO qubits.   
Electrons are vertically confined in a 5-nm-thick silicon well surrounded by isotopically natural Si$_{0.3}$Ge$_{0.7}$ barriers.  
Silicon in the well is isotopically enhanced to increase the proportion of nuclear-spin-zero isotopes, with a residual 800-ppm content of \isotope[29]{Si}.  
Lateral confinement is achieved with the electrostatic potentials induced by tuned voltages on lithographically patterned metallic gates [\reffig{fig:device}(a, b)].
Voltages on plunger gates (P1 through P6) accumulate a single electron in each dot, while exchange gates (X1 through X5) control the interaction strength between neighboring electrons. 
Additional tunnel (T1, T6) and bath gates (B1-B3) control coupling to electron reservoirs \cite{Borselli2011}.
We measure the device charge configuration using two integrated dot charge sensors [\reffig{fig:device}(a)] which we refer to as M1 and M2 and whose barrier gates are Z1, Z2 and Z3, Z4 respectively. 

\begin{figure*}[ht]
	\centering
	\includegraphics[width=1.00\textwidth]{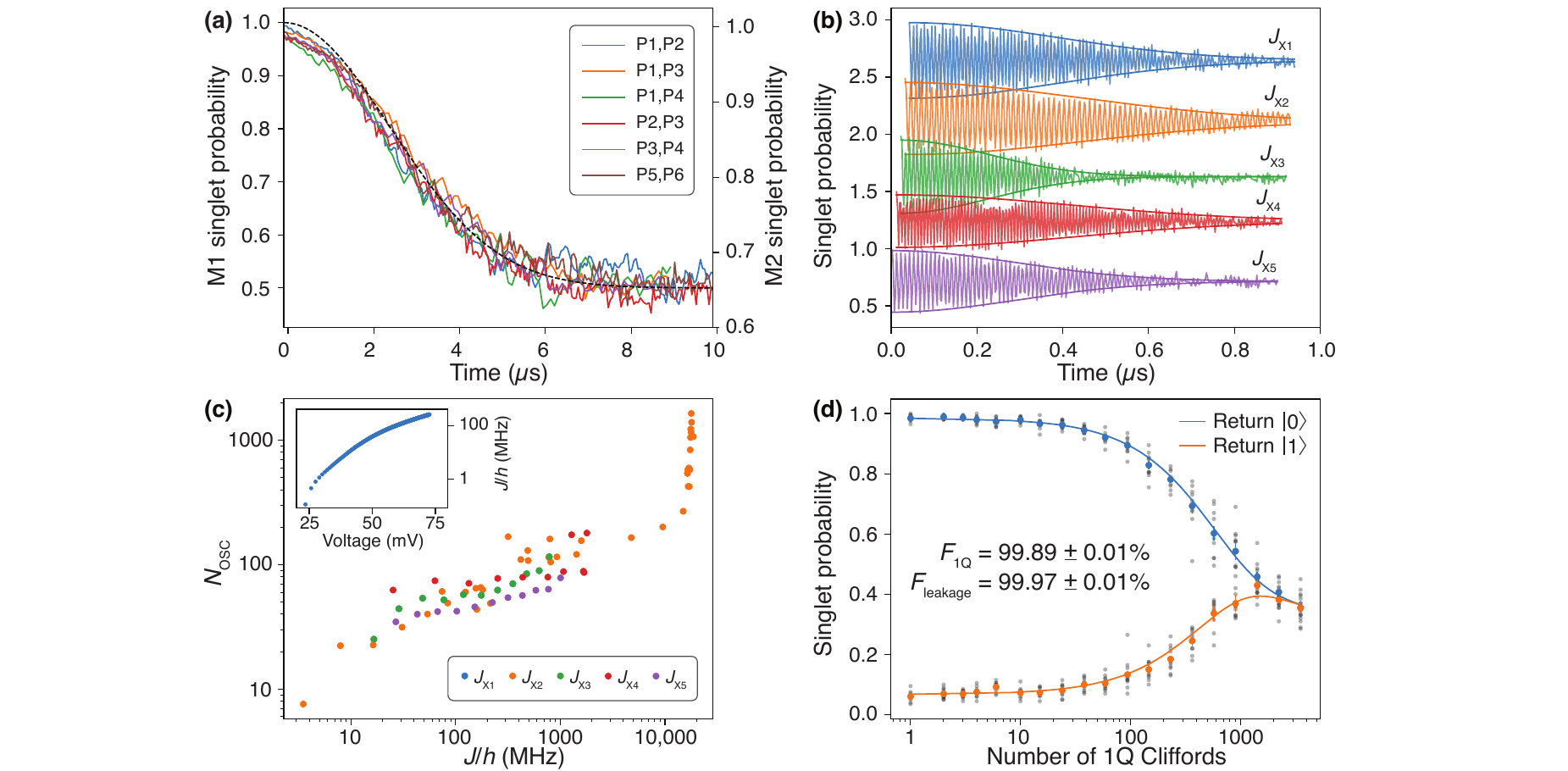}
	\caption{
		Device performance metrics. 
		(a) Magnetic dephasing of a spin-singlet at the idle position (where $J \approx 0$), prepared on different dot pairs.
		The 1/$e$ point of the gaussian decay envelope defines $t=\ttwostar$ and is a simple metric for characterizing the impact of substrate nuclear magnetic noise on qubit performance.  
		We plot a $\ttwostar = 3.5$~$\mu$s envelope as a visual guide in dashed black.
		Most pairs are prepared and measured with high fidelity on the M1 side; the P5/P6 pair uses M2 and has lower contrast.
		(b) Charge noise impact on exchange oscillations for each exchange axis.
		This measurement is analogous to the one for $\ttwostar$ except measured at $J / h \approx 100$~MHz such that fluctuations in the exchange energy due to charge noise are the dominant source of decoherence.
		We parameterize the 1/$e$ gaussian decay point in terms of the number of coherent oscillations, \nosc, that occur in that time.
		Each successive curve is offset on the $y$-axis by 0.5 and on the $x$-axis by 10~ns.
		(c) \nosc\ as a function of $J$.  
		Due to the sub-exponential behavior of exchange along the symmetric axis ~\cite{reed2016} (inset), we observe that \nosc~increases with $J$. 
		At $J / h \geq$ 17~GHz, exchange asymptotes due to the flattening of the tunnel barrier to a value related to the double-dot orbital energy \cite{dial2013}.
		In this limit, where voltage throws on numerous gates exceed $\pm 200$~mV, $dJ/dV$ decreases significantly and \nosc~rapidly increases to an observed maximum $>1600$.
		(d) Single-qubit performance for the P1, P2, P3 qubit with M1 readout. The ``blind'' randomized benchmarking measurement \cite{andrews2019} yields an average single qubit gate error of $(1.1 \pm 0.1) \expn{-3}$ and leakage error of $(3 \pm 1) \expn{-4}$.
		See \refapp{app:sources} for more detailed discussion of this figure.
	}
	\label{fig:nosc_t2star_brb}
\end{figure*}

Our two qubits are each encoded into three physical spins as described in Ref.~\onlinecite{andrews2019}.  
The qubit states, which have total (three-spin) angular momentum $S=1/2$, are defined by the angular momentum of the two spins (1 and 2) closest to the measurement reservoir.
The encoded $\ket{0}$ state corresponds to two-spin singlet ($S_{12}=0$) and encoded $\ket{1}$ corresponds to two-spin triplet ($S_{12}=1$).  
Qubit states are measured by mapping their spin state to charge configuration with Pauli spin blockade \cite{blumoff2022, burkard2021}.
A similar physical mechanism is used to initialize qubits into singlet states \cite{blumoff2022}.  
We prepare and measure spin states on the outermost dot pairs (P1/P2 and P5/P6), with the X3 gate connecting the innermost electron spins in dots P3 and P4.
The fidelity of state preparation and measurement (SPAM) operations are in part limited by valley excited-state energies of the dots, which we measure using photon-assisted tunneling to be 70 $\mu$eV for dot P1 and 14 $\mu$eV for dot P6.
The relatively poor valley splitting on P6 is consistent with observations of reduced SPAM fidelities on that side of the device and is why we later prioritize M1 measurement when possible (see \refapp{app:bilateral}).

\begin{figure*}
	\includegraphics[width=1.00\textwidth]{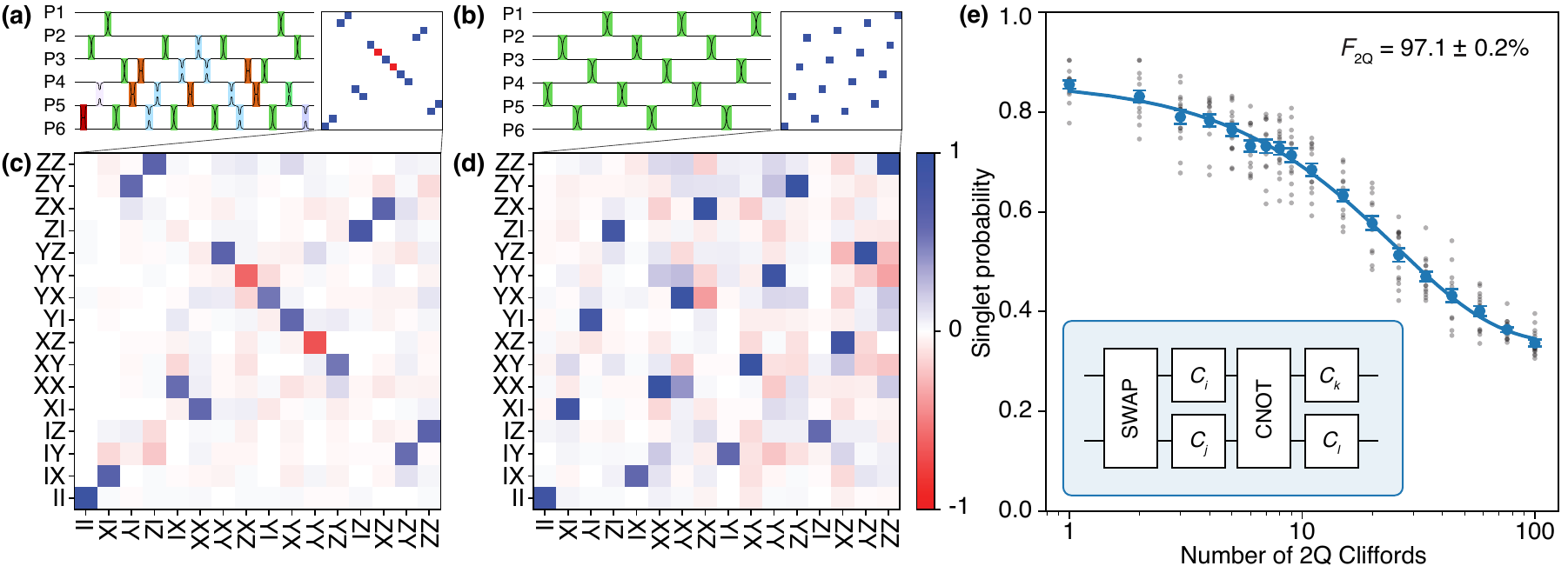}
	\caption{
		Two-qubit process tomography and randomized benchmarking.
		(a-b) Exchange pulse diagrams and ideal process matrices of FW-CNOT and SWAP gates.
		The shading of exchange pulse boxes is proportional to the pulse exchange angle.
		(c-d) Maximum-likelihood estimates of measured quantum process matrices of FW-CNOT and SWAP gates.
		These data depend on joint measurement, so the relatively low M2 SPAM fidelity reduces contrast.
		(e) Two-qubit randomized benchmarking with \tpulse = 10~ns, \tidle = 5~ns, and applied magnetic field $B$ = 2.1~mT.
		Two-qubit Clifford gates are compiled using FW-CNOT, SWAP, and single-qubit Clifford gates $C_x$ (inset). 
	}
	\label{fig:tomo_2qrb}
\end{figure*}

The qubits are controlled by sequentially applying exchange rotations between neighboring spins [\reffig{fig:device}(c)].
A single exchange pulse is actuated by voltage-modulating gate X$n$ (and neighboring P$n$ and P$(n+1)$ gates for capacitive compensation to achieve symmetric operation~\cite{reed2016}) and drives a rotation about the corresponding exchange axis $J_{\text{X}n}$.
Exchange coupling energies between dots are kept low ($J_{\text{X}n}/h<10$~kHz, $h$ being Planck's constant) when in the ``idle'' configuration but pulsed to a large value ($J_{\text{X}n}/h \approx$ MHz to GHz) to drive rotations.
Moving between these regimes requires control pulse amplitudes of $\sim\!\!\!100$ mV; the approximately exponential relationship between $J$ and voltage \cite{reed2016} strongly suppresses control crosstalk to adjacent idling electrons.
Sequences are built up with pulses of fixed duration \tpulse, with pulse-to-pulse spacing \tidle.
Our pulse calibration routine \cite{andrews2019} yields a continuous mapping $\theta_{n}(\vec{V}_{n})=\int dt J_{\text{X}n}[\vec{V}_{n}(t)]/\hbar$ as a function of voltage throw $\vec{V}_{n}$ along all five of the symmetric axes~\cite{reed2016}, from which we generate voltage pulse amplitudes for all required gate sequence angles.
We provide control performance metrics as in \cite{reed2016, andrews2019} in \reffig{fig:nosc_t2star_brb} and further elaborate in \refapp{app:sources}.

Our demonstration of two-qubit control of this improved device is a direct extension of earlier single-qubit work \cite{andrews2019}.
This is in part because the EO modality, unlike most other qubit systems \cite{blais2021, monroe2021}, does not require a new physical mechanism to generate entanglement between encoded qubits.
We diagram the FW-CNOT and SWAP pulse sequences in \reffig{fig:tomo_2qrb}(a) and (b), which are respectively made of 28 and 15 exchange operations spanning five control axes.  
The construction of these gates reflects our physical device connectivity and choice to initialize the outer two pairs of spins.
Pulses are applied sequentially in order to limit crosstalk, but future larger devices may support simultaneous operation.
We first characterize both operations using quantum process tomography (QPT)~\cite{Merkel2013} as initial confirmation that we are performing the intended encoded gates, shown in \reffig{fig:tomo_2qrb}(c) and (d).
We see qualitative agreement with expected tomograms, but, due to the weakness of our QPT inversion method to leakage and SPAM errors, and in particular the poor M2 SPAM in the present device, it is difficult to extract meaningful quantitative results from this protocol.
This could in principle be mitigated with increased averaging and self-consistent gateset tomography~\cite{Nielsen2020, madzik2022, xue2022}, but we do not explore that here.

\begin{figure*}[!t]
	\includegraphics[width=1.00\textwidth]{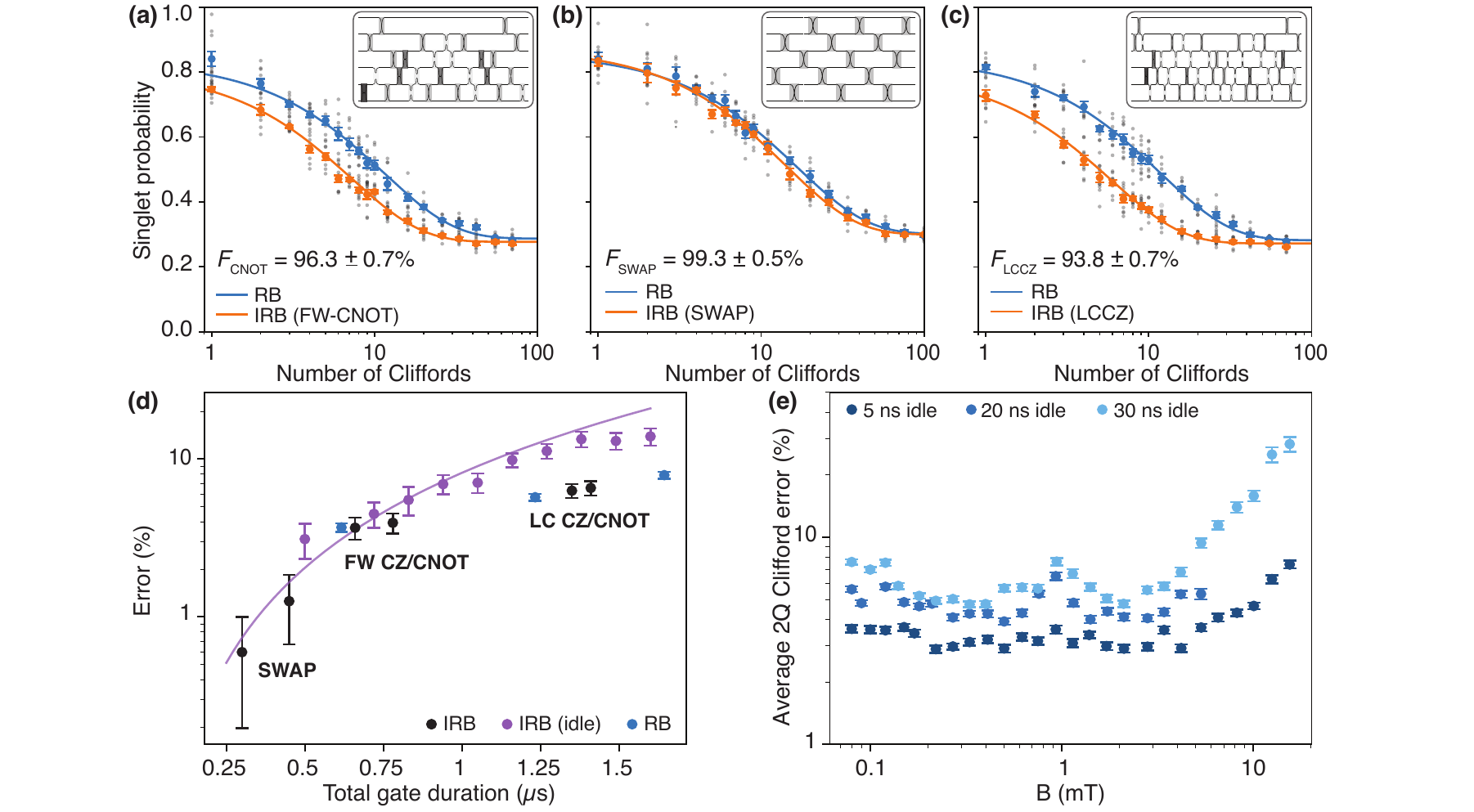}
	\caption{
		Interleaved two-qubit benchmarking and error trends. 
		(a-c) Benchmarking of SWAP, FW-CNOT, and LCCZ at $B=0$, with inset exchange pulse diagrams.
		Here \tidle = 20~ns for FW-CNOT and LCCZ but \tidle = 10~ns for SWAP; \tpulse = 10~ns for all.
		(d) Error as a function of gate duration for selected gates (black) (see \refapp{app:sources}), for various interleaved idle periods (purple) and for two-qubit Cliffords (blue).
		Idle error increases with gate duration, closely following a theoretical estimate given by $(\tidle/\ttwostar)^2$ (purple curve).
		Some gates show a significant deviation below this curve, indicating that they have some built-in magnetic noise insensitivity.
		(e) Average 2-qubit Clifford gate error as a function of $B$ and pulse idle time.
		$B$ is oriented in-plane and perpendicular to the dot array.
		We see consistent improvement for lower \tidle, eventually limited by the available bandwidth of the signal chain (not shown). 
		As $B$ increases, we first observe an improvement in fidelity above 200~$\mu$T, consistent with the suppression of transverse hyperfine magnetic gradients. 
		The fidelity decreases for $B>3$~mT due to induced paramagnetic gradients. 
	}
	\label{fig:irb}
\end{figure*}

Randomized benchmarking (RB) is our preferred method of characterizing gate performance since it is fast, simple, relatively insensitive to SPAM error, and requires measurement of only one qubit ~\cite{knill2008,emerson2005}. 
In RB, a randomly-selected sequence of gates that compile to the identity is chosen from a discrete group of qubit operations, typically the Clifford group.
This choice depolarizes noise in the encoded subspace, allowing gate performance to be inferred by sweeping the sequence length and fitting return probability to an exponential decay.
We generate two-qubit Cliffords via standard compilation rules ~\cite{koenig2014} using the FW-CNOT entangling gate, the aforementioned SWAP gate, and single-qubit Cliffords.
Of the 11520 two-qubit Clifford gates, 90\% include a CNOT in our composition, 50\% include a SWAP, and each has an average of $3.1$ single-qubit Clifford gates; each operation thus contains 41.1 exchange pulses on average. 
As shown in \reffig{fig:tomo_2qrb}(e), we find an average 2-qubit Clifford fidelity of \twoRBF, far better than suggested by the SPAM-afflicted QPT.
The fidelity we indicate would be the entanglement fidelity of average Clifford gates for the encoded state without leakage,
but our technique does not correctly attribute leakage error versus encoded error; an extension to the blind RB protocol ~\cite{andrews2019} for two-qubit RB is an ongoing effort.  
This complexity, as well as the leakage calculation predicting an asymptote of 17/60, are discussed in \refapp{app:bilateral}.

We also measure the performance of individual Clifford gates via interleaved randomized benchmarking~(IRB)~\cite{magesan2012}.
In this protocol, each random gate is interleaved with an identical copy of the operation in question and the resulting decay is compared to a reference RB decay to infer the fidelity of the chosen operation.
We perform IRB for FW-CNOT, SWAP, and LCCZ in \reffig{fig:irb}(a-c), finding a fidelity of \cnotF, \swapF, and \lcczF, respectively.
The SWAP error is more than five times lower than that of CNOT; the existence of this high-fidelity native SWAP gate in the EO modality, in contrast to most other qubit technologies, may enable new device topologies.
The estimated error due to magnetic noise for these sequences is found via numeric simulation and is consistent with the measured data, up to uncertainties in the actual magnetic field witnessed at the quantum dot (see \refapp{app:sources}).  
These same simulations, elaborated in Ref.~\cite{andrews2019}, indicate that the impact of $1/f$ charge noise, quantified by exchange oscillations, account for less than 6\% of the observed error.  

We expand on this magnetic noise dependence in \reffig{fig:irb}(d) by plotting gate fidelity as a function of total gate duration.
We also plot the interleaved fidelity of idle operations of increasing duration $t$, whose estimated theoretical error is $\epsilon=\left(t / \ttwostar \right)^2$, where $\ttwostar \approx 3.5 \mu s$ is the $1/e$ decay time of an idle singlet state (see \refapp{app:sources})~\cite{kerckhoff2021}.   
Unlike charge noise, magnetic noise acts during both pulsing and idling~\cite{ladd2012}, and so dephasing occurs continuously during evolution, leading to a total error scaling as $(t_{\rm gate}/\ttwostar)^2$. 
The prefactor of that scaling (generally $\leq 1$), however, depends on how much the sequence permutes spins and decouples magnetic noise~\cite{kerckhoff2021}. 
For this reason, most sequences in \reffig{fig:irb}(d) fall beneath the $\epsilon=\left(t_{\rm gate} / \ttwostar \right)^2$ error level of ``doing nothing'', consistent with their errors being dominated by magnetic noise.
As an example, the LCCZ operation shown here has roughly half the infidelity of an equal-duration idle.

Finally, we investigate the dependence of gate fidelity on both global magnetic field strength and sequence duration.
Here, the magnetic field is aligned in-plane with the device substrate and approximately perpendicular to the linear dot array.
Though the EO encoding is explicitly immune to static and fluctuating global field (it is a ``decoherence free subsystem" in this regard), such global fields impact the magnitude of local magnetic field gradients.
In \reffig{fig:irb}(e) we plot average two-qubit Clifford fidelity measured using RB as a function of both applied field and \tidle, and find that error initially decreases by nearly a factor of two with increasing field strength.
We understand this improvement as the suppression of nuclear magnetic fluctuations transverse to the field direction \cite{kerckhoff2021}.
At fields $>3$~mT, error increases again due to a combination of spin-orbit effects and Meissner screening effects from superconducting parts of the gate-stack (see \refapp{app:sources}).
This effect is stronger in earlier devices employing aluminum metal~\cite{kerckhoff2021}, but is much weaker in the present device due to the non- or weakly-superconducting TiN gates used in the SLEDGE process~\cite{haha2021}.
From this plot we may infer that the IRB shown in \reffig{fig:irb} might be further optimized by operating at finite magnetic field, data which is the subject of future work.

In this report, we have demonstrated universal logic operations with two EO qubits in silicon, including single-qubit gates, CNOT, CZ, and SWAP.
We have also demonstrated LCCZ, which is expected to reduce leakage spreading in future quantum processors.
The dominant error source of all entangling gates is hyperfine magnetic noise.
Combined with the other advantages of silicon spin qubits, the EO modality constitutes a compelling vision for achieving FT.
Future work toward this goal includes detailed error budgeting and validation of leakage-sensitive two-qubit gate characterization, multiple back-end metal layers using SLEDGE to enable larger devices, increased isotopic enhancement to reduce magnetic noise, and improved signal conditioning for faster operation.

\textbf{Acknowledgements}. 
We thank John B. Carpenter for assistance with all figures, and acknowledge significant technical contributions from 
Edwin Acuna, Ed Chen, Lisa Edge, Mark Gyure, Andy Hunter, Cody Jones, Brett Maune, Seth Merkel, Ivan Milosavljevic, Emily Pritchett, Richard Ross, Adele Schmitz, Chris Schnaible, Bo Sun, Roland Velunta, Stephen Wandzura, and Parker Williams. 

\newpage
\appendix

\section{Theory of multiqubit exchange-only pulse sequences}
\label{app:theory}
In this section we explain the structure of the entangling pulse sequences we demonstrate, including the purpose and construction of the LCCZ gate.

These entangling gates act on encoded subspaces of multiple spins which are best described according to angular momentum quantum numbers.  
A single spin, say spin 1, has $S_1=1/2$, and projection $m_1=\pm 1/2$.  
For two spins, labelled 1 and 2, the total angular momentum $S_{12}$ can be $S_{12}=0$, the singlet state for which $m_{12}=0$, or $S_{12}=1$, the triplet of states for which $m_{12}=-1,0,$ or $1$.  
We notate these four states as $\ket{S_{12};m_{12}}$.  
The smallest universally-controllable exchange-only qubit requires adding a third spin, for which the algebra of angular momentum provides the quantum number $S_{123}$, which can be $1/2$ either by adding the single $S_3=1/2$ spin to the $S_{12}=0$ singlet, or by adding $S_3$ to the $S_{12}=1$ triplet.  
These two choices for $S_{123}=1/2$ are our encoded state, and the quadruplet of states with $S_{123}=3/2$, which also results from adding $S_3$ to the $S_{12}=1$ triplet, are leaked states.  
We also have a total spin projection $m_{123}=-S_{123},-S_{123}+1,\ldots,S_{123}$.  
In this section we thus notate these states as $\ket{S_{12},S_{123};m_{123}}$ and our qubit states are $\ket{0,1/2;m_{123}}$ and $\ket{1,1/2;m_{123}}$.  
The $m_{123}$ degree of freedom is called the gauge, and it is unaffected by exchange operation between any of the three spins.  
The Pauli spin blockade process on dots 1 and 2 provides initialization and measurement only of the $S_{12}$ quantum number and has no impact on $m_{123}$ \cite{blumoff2022}.   
Singlet measurements on spins 1 and 2 therefore measure the probability of $\ket{0,1/2;m_{123}}$ independent of $m_{123}$, and provide no information to distinguish encoded triplet $\ket{1,1/2;m_{123}}$ states from leaked states $\ket{1,3/2;m_{123}}$.  
To understand single-qubit gates in the single-qubit system, see Ref.~\onlinecite{andrews2019}.

\begin{figure}
	\centering
	\includegraphics[width=\columnwidth]{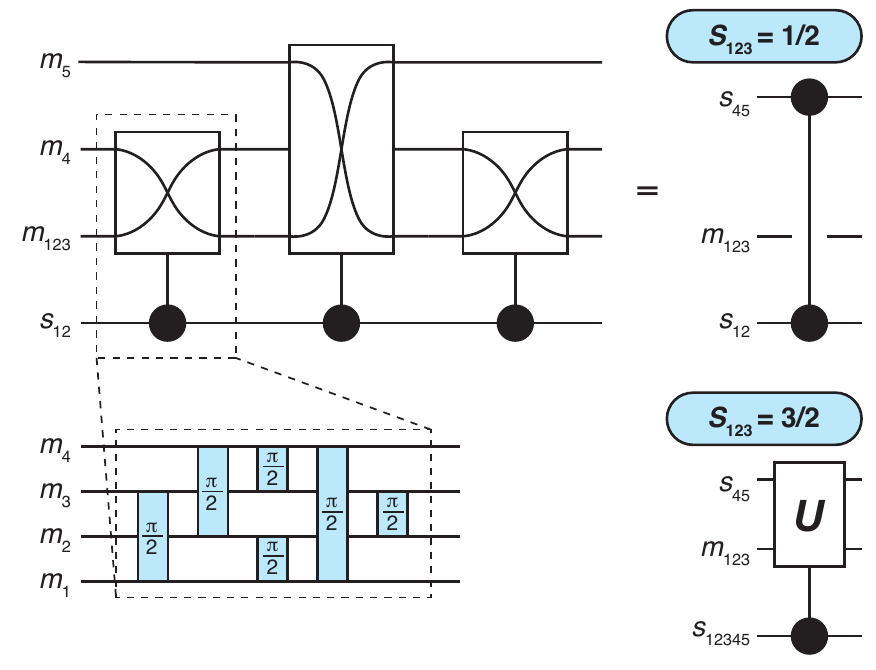}
	\caption{Mathematical construction of the FWCZ.  
	A primitive subsequence coupling all pairs of four spins with $\pi$/2 pulses (spin-$\sqrt{\text{SWAP}}$) gates is shown in the dashed box, which, in the $S_{123}=1/2$ (encoded) subsystem, is controlled by the $S_{12}$ quantum number of the $\ket{S_{12},S_{123};m_{123}}$ qubit.  
	The primitive is identity for $S_{12}=0$ and swaps $m_{123}$ with a fourth spin if $S_{12}=1$.  
	Three uses with alternating choice of the fourth spin completes a controlled-$Z$ between singlet-triplet subsystems on $S_{12}$ and $S_{45}$. 
	If $S_{123}=3/2$ (leakage subsystem), the gate applies an $S_{12345}$-dependent unitary $U$ to $S_{45}$, with $U^2=1$.  
	A full-pulse construction of this FWCZ including the additional $\pi$-pulses (spin-SWAPs) to adapt to a linear nearest neighbor layout is shown in Table~\ref{tab:gate-sequences}.}
	\label{fig:FWCZconstruction}
\end{figure}
When considering operations on two qubits on dots 1-6, the gauges of the two qubits, $m_{123}$ and $m_{456}$, become important.  
Notating the total angular momentum of all six spins as $S$ rather than $S_{123456}$ and their projection as $m$ instead of $m_{123456}$ for brevity, the two $S_{123}=S_{456}=1/2$ qubits may combine into an $S=0$ subspace and an $S=1$ subspace.  
While exchange conserves $m$, still respecting gauge freedom, its action for interqubit operations does depend on $S$, which in turn is set by the relative value and phase of $m_{123}$ and $m_{456}$.
The gauge freedom, which can be safely ignored for single-qubit gates, must therefore be carefully considered with two-qubit operations.

\begin{figure}
	\centering
	\includegraphics[width=\columnwidth]{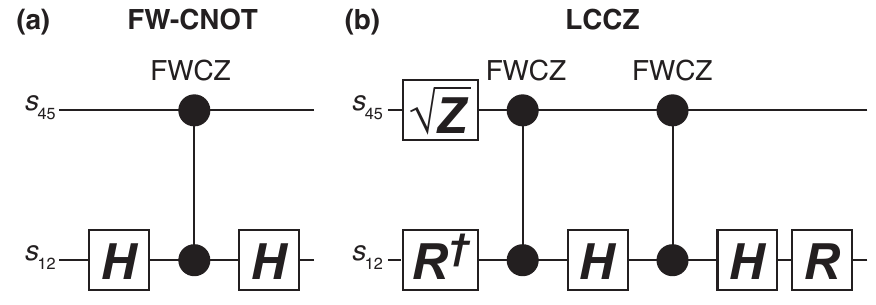}
	\caption{(a) FW-CNOT is made from FWCZ (Fig.~\ref{fig:FWCZconstruction}) in the standard way: two Hadamard gates notated $H$ as in Ref.~\onlinecite{andrews2019} and some compiling. 
	(b) The LCCZ is made from two FWCZ interspersed with single-qubit gates, and some compiling.  
	Here $R=(I+iZ)/\sqrt{2}$, for identity $I$ and Pauli $Z$.  
	Full-pulse constructions are shown in Table~\ref{tab:gate-sequences}.}
	\label{fig:LCCZconstruction}
\end{figure}

In 2000, Ref.~\onlinecite{divincenzo2000} provided a 19-pulse entangling gate sequence between two encoded qubits that required the total angular momentum of all six spins to be $S=1$, which was imagined to be accomplished by polarizing the gauge of each qubit.  
The small spin-orbit effect and long spin relaxation times in silicon leave few good hardware choices for such polarization in our system.  
A decade later, the Fong-Wandzura (FW) construction \cite{fong2011} was discovered by computational search employing a genetic algorithm.  This construction is a gauge-independent CNOT sequence, meaning it correctly performs the same CNOT sequence on the $S_{12}$ and $S_{45}$ degrees of freedom for both $S=0$ and $S=1$.  Gauge independence allows each triple-dot EO qubit to be initialized as a pair of spin-singlet states and an unpolarized spin, as we do in the present demonstration.  

Half a decade later, Ref.~\onlinecite{zeuch2016} showed that the FW sequence is in fact composed of three repetitions of a shorter primitive composite sequence acting on four spins, shown in \reffig{fig:FWCZconstruction}.  
This primitive sequence is a quasi-Fredkin (controlled-SWAP) gate, swapping the gauge $m_{123}$ with $m_4$ only if $S_{12}=1$, but not if $S_{12}=0$.  
Applying this quasi-Fredkin gate to one qubit on spins 1,2, and 3 and alternatingly on spins 4, 5, and then 4 again, a $S_{12}=0$ condition will apply identity three times, while an $S_{12}=1$ condition will swap spins 4 and 5, leaving $m_{123}$ in its initial state (regardless of what that initial state is).   
These three uses provide a Fredkin gate with a three-spin EO qubit as control and two spins as target.  
If those two spins are the singlet-triplet pair of an EO-qubit, this controlled-SWAP becomes an encoded CZ, the FWCZ, with no gauge dependence.  

\begin{figure*}[t]
	\includegraphics[width=1\textwidth]{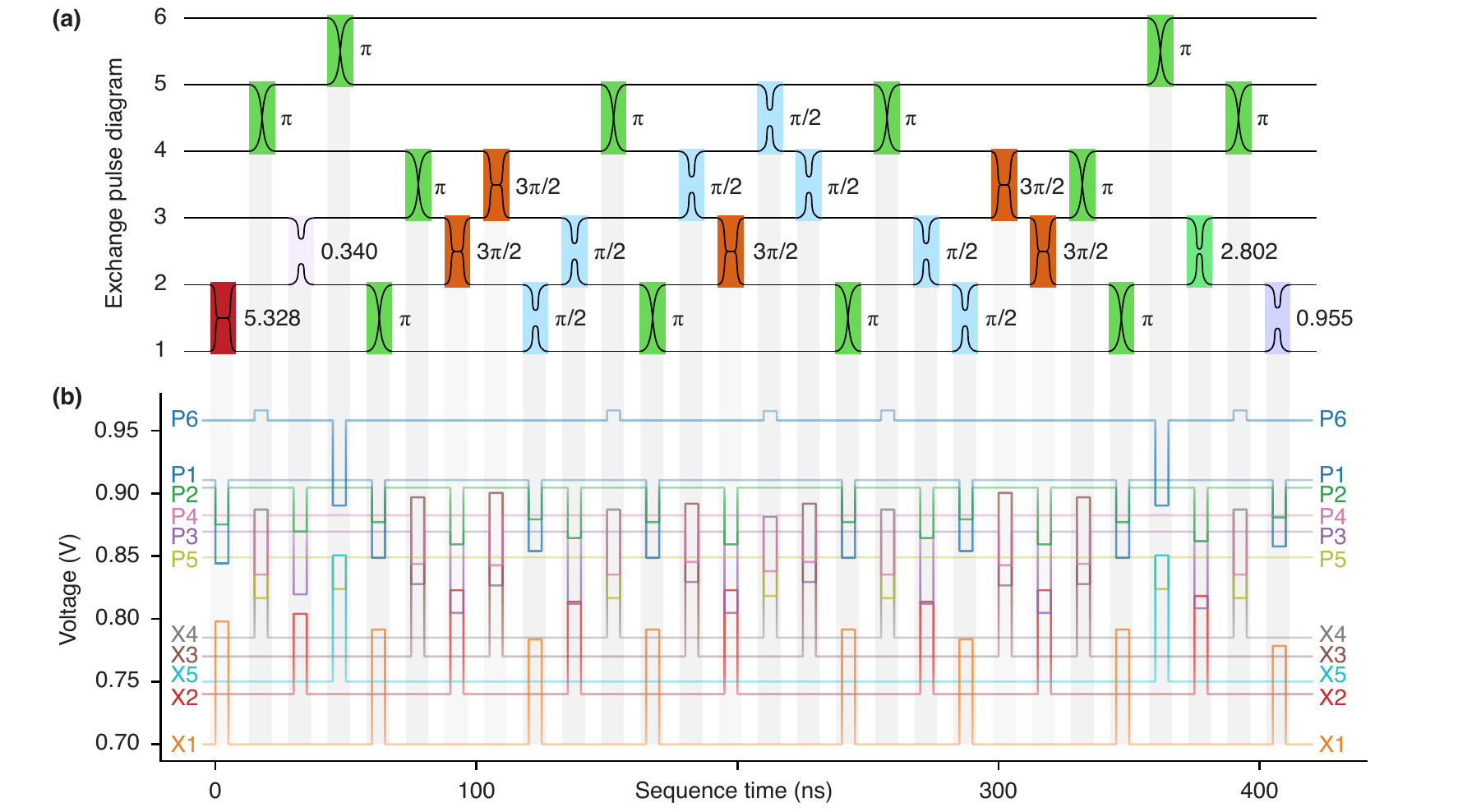}
	\caption{A Fong-Wandzura CNOT and its corresponding voltage control waveform. 
	Pulse angles (a) defined on exchange axes are translated into rectangular voltage pulses (b) specified at the gates, with varying amplitudes along the corresponding symmetric axes.  
	The voltages waveforms shown here are the exact voltage waveforms used in the RB experiment of \reffig{fig:tomo_2qrb}(c).
	}
	\label{fig:app-voltage-waveform}
\end{figure*}

Compiling the primitive sequences together, one arrives at an entangling gate using 12 $\pi/2$ exchange pulses (i.e. spin $\sqrt{\text{SWAP}}$ gates) on just five fully-connected spins; assuring a controlled-$Z$ adds two more $\pi/2$ pulses, and adapting to the linear, nearest-neighbor coupled layout we employ here with the measured singlet-triplet pairs on the ends of the array, the FWCZ ends up using all 6 spins with an additional 12 $\pi$ pulses (SWAPs on spins) for a total of 26 pulses, shown in Table~\ref{tab:gate-sequences}.  
A CZ converts to CNOT via the construction shown in Fig.~\ref{fig:LCCZconstruction}(a), adding two more pulses.
To highlight the physical implementation of such sequences, we present an illustrative example in \reffig{fig:app-voltage-waveform} of the 28-pulse FW-CNOT sequence translated into experimentally-accurate voltage waveforms required for device control.
The sequences are compiled sequentially, so that no two pulses occur simultaneously.

Unfortunately, the quasi-Fredkin gate on four spins discussed above has an undesired feature: if the EO qubit is in its $S_{123}=3/2$ leaked state, then the gate applies a phase flip to the $S_{1234}=2$ states relative to the $S_{1234}=1$ states. Mathematically, no exchange-only four-spin sequence can avoid this problem.  
As a result, when $S_{123}=3/2$ and we apply this primitive operation as described for the FWCZ, then when $S_{12345}=1/2$ and $S_{45}=1$, the resulting unitary provides a phase flip, and when $S_{12345}=3/2$, the resulting unitary applies a $\pi$ rotation about an axis tipped an angle $\tan^{-1}(3\sqrt{15}/11)$ from the Bloch-sphere $z$-axis to the singlet-triplet qubit defined by $S_{45}$.  
These operations in general will leak the EO qubit including spins 4 and 5, and as a result, even when applying the FW gate perfectly, leakage will have spread from one qubit to the next.  
(Notably, leakage spreads only from the $\ket{S_{12},S_{123};m_{123}}$ qubit to the $\ket{S_{45},S_{456};m_{456}}$ qubit; if the $\ket{S_{45},S_{456};m_{456}}$ is leaked into $\ket{1,3/2;m_{456}}$ then the gate behaves as intended for $S_{45}=1$ regardless of $S_{456}$).  
This leakage spreading could be highly detrimental to fault-tolerance, as we are unable to detect leakage directly in general and most quantum error correcting codes are ill-equipped to correct leakage even when it is detected.  

The goal of the LCCZ gate is to avoid this leakage spreading when applying encoded CZ gates.  
The key insights to the LCCZ gate are that the unwanted leakage-induced phase flip or $\pi$ rotation on the $S_{45}$ qubit are both square-roots of an identity operation, and all single-qubit operations are identity on leakage spaces.  
Therefore, if we apply two FWCZ sequences with a single-qubit gate on dots 1-3 in between, then on the encoded subspaces we have achieved some controlled-$\pi$-rotation gate, where the $\pi$-rotation angle depends on the choice of single-qubit operation and can be converted back to Pauli operators $Z$ or $X$ with single-qubit corrections.  
In particular, we use the construction shown in Fig.~\ref{fig:LCCZconstruction}(b).
The operator $\sqrt{Z}$ (often called $S$, but not to be confused with total spin) on $\ket{S_{45};m_{12}}$ is simply another $\pi/2$ pulse on these spins, and the single-qubit rotations $R$, $R^\dag$, and $H$ may be readily derived as exchange sequences similar to those in Ref.~\onlinecite{andrews2019}.  
If the CZ in this construction is the FWCZ and the $S_{12}$ qubit is leaky (i.e. $S_{123}=3/2$), then all single-qubit gates have no action and the two FWCZ gates combine into identity, leaving behind only the correctable single-qubit-encoded $\sqrt{Z}$ on the $S_{45}$ pair, (which is also $\sqrt{Z}$ on a $\ket{S_{45},S_{456};m_{456}}$ qubit).  
After some compiling, the resulting sequence is 46 pulses for an LC-CNOT and 44 pulses for an LCCZ; the compiled sequences are shown in Table~\ref{tab:gate-sequences}. 

The final sequence we demonstrate is SWAP, which is not entangling, but is nonetheless a critical two-qubit gate for moving data and randomized benchmarking. 
In many qubit modalities, the SWAP is more complex than CNOT, as a typical construction uses three CNOT sequences.  
However, for exchange-only qubits, SWAP is the one transversal operation of the underlying spins.  
If our spins were fully connected, three $\pi$ pulses to enact spin swaps would suffice.  
The 15 pulse sequence shown in Table~\ref{tab:gate-sequences} performs the permutations needed to move spin information in linear, nearest-neighbor coupled architecture with the three-spin structure reflected about the center.  
The timing of these spin-swap operations is chosen so that each spin spends approximately the same amount of time in each dot.  
A sequence in which this is done exactly effectively decouples low-frequency magnetic noise~\cite{west2012, sun2022}; the 15-pulse SWAP shown only partially completes such a permutative dynamical decoupling operation.  

\begin{table*}
    \caption{Two-qubit gate sequences.}
    \label{tab:gate-sequences}
    \includegraphics[width=0.85\textwidth]{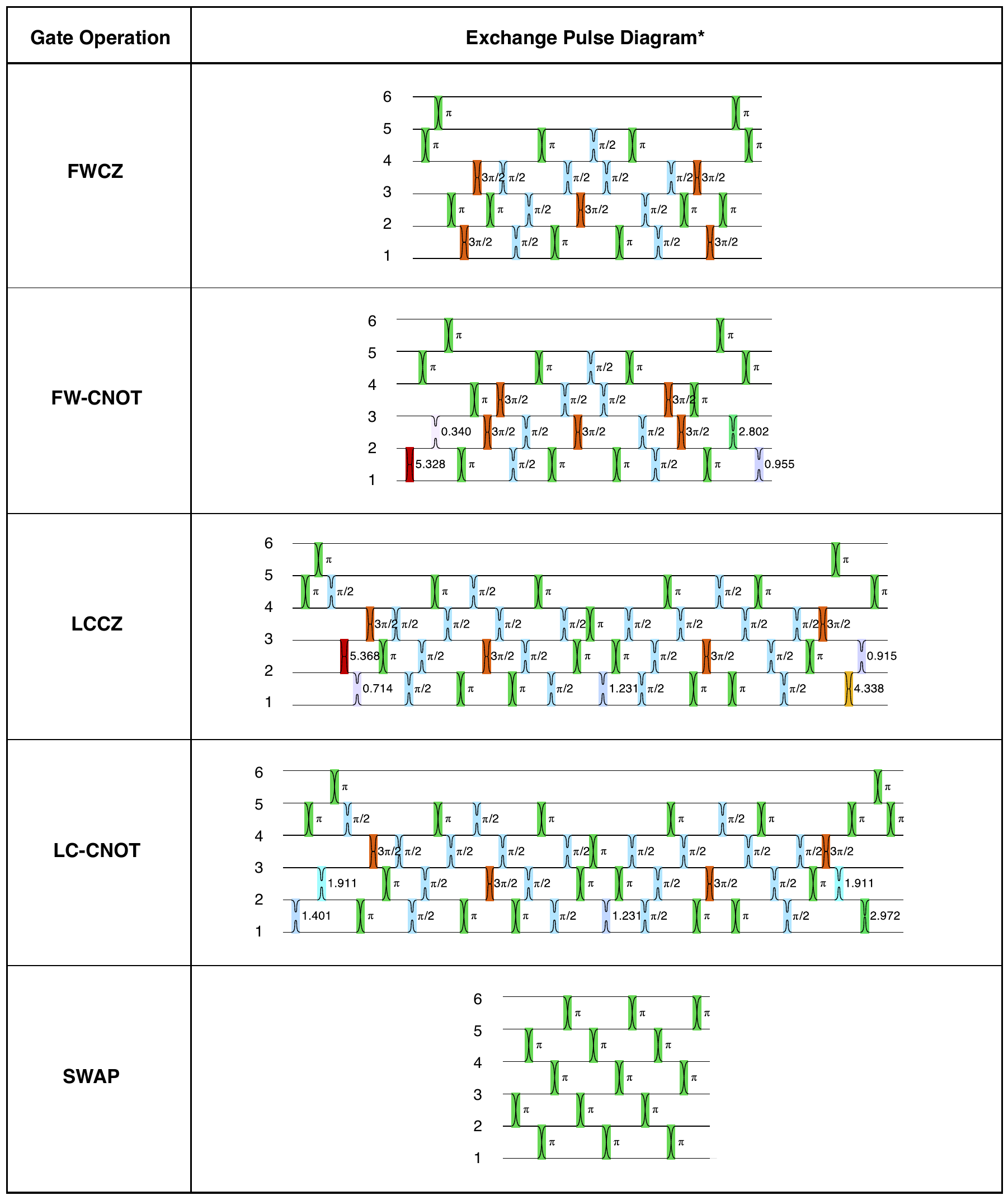}
    \newline
    \footnotesize\textsuperscript{*}Angles presented here in decimal format are shorthand for the following expressions:  $\cos^{-1}\left(2\sqrt{2}/ 3\right) \rightarrow 0.340$, $\cos^{-1}\left[\left(2 - \sqrt{2} + 2 ^{3/4}\right)/3\right] \rightarrow 0.714$, $\cos^{-1}\left[\left(-1 - 2\sqrt{2}\right)/ 3\right] \rightarrow 0.915$, $\cos^{-1}\left(1 / \sqrt{3} \right) \rightarrow 0.955$,  $\cos^{-1}\left(1 / 3 \right) \rightarrow 1.231$, $\cos^{-1}\left( 1 / \sqrt{3} - 1 / \sqrt{6} \right) \rightarrow 1.401$, $\cos^{-1}\left(-1/3 \right) \rightarrow 1.911$, $\cos^{-1}\left( -2 \sqrt{2} / 3 \right) \rightarrow 2.802$, $\cos^{-1}\left( \sqrt{ 1 / 2 + \sqrt{2} / 3} \right) \rightarrow 2.972$, $2\pi - \cos^{-1}\left[\left(2 - \sqrt{2} + 2 ^{3/4}\right)/3\right] \rightarrow 4.338$, $2\pi - \cos^{-1}\left( 1 / \sqrt{3} \right) \rightarrow 5.328$ and $2\pi - \cos^{-1}\left[\left(-1 -2 \sqrt{2} \right) / 3 \right] \rightarrow 5.368$.
\end{table*}

\section{Two-qubit benchmarking with bilateral readout and inversion rotations}
\label{app:bilateral}

\begin{figure}[htp]
    \includegraphics[width=\columnwidth]{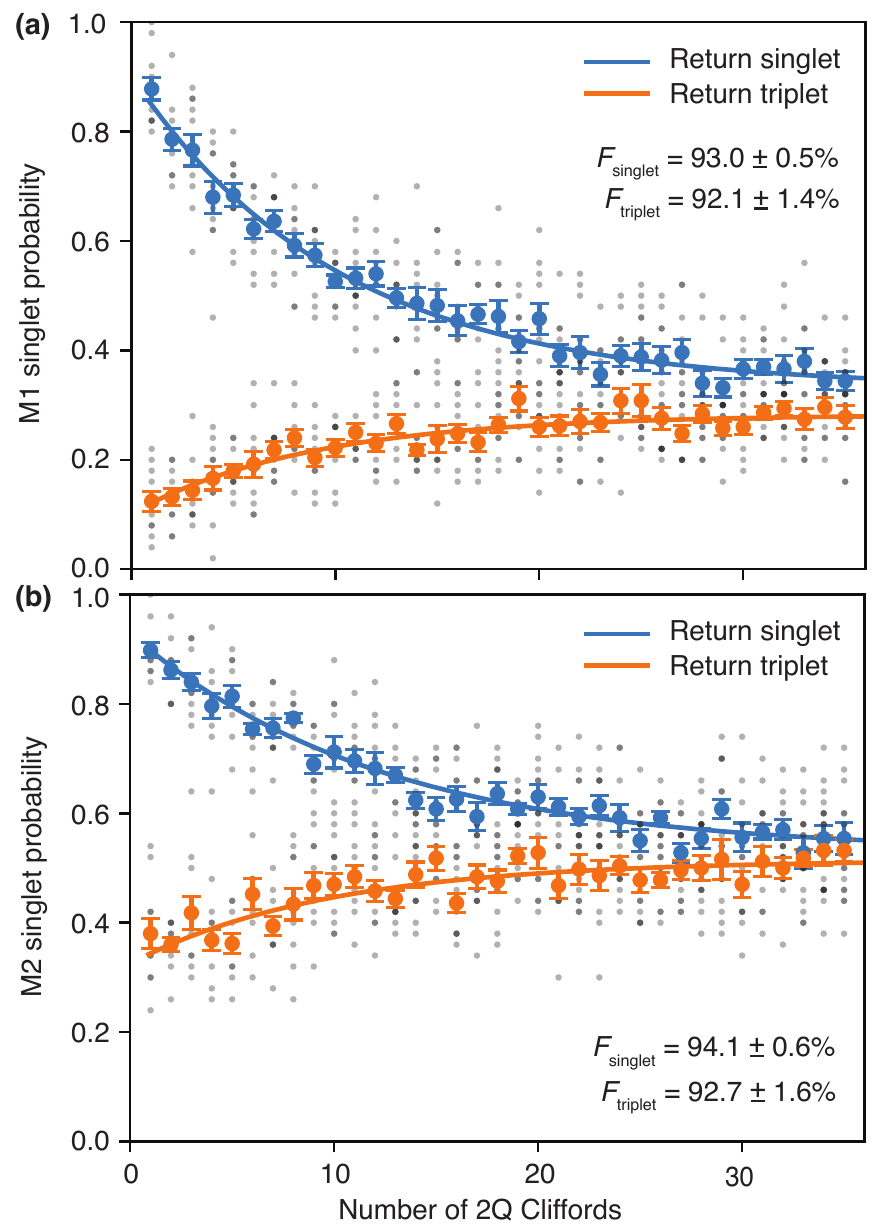}
    \caption{Two-qubit randomized benchmarking with two-sided readout, both with and without $X$-gate pre-rotations.  
    Cliffords are compiled with the FW-CNOT entangling gate.  
    We observe worse performance here compared to the RB datasets in the main text as we operate here with \tidle = 20 ns, compared to \tidle=5 ns in \reffig{fig:tomo_2qrb}(c), but consistent with performance observed in \reffig{fig:irb}(e).
    }
    \label{fig:app-2qrb-m1h-m2h}
\end{figure}

Though all RB results in the main text are generated via one-sided readout only, we present here benchmarking results with readout from both sides.
As stated in the text, small valley splitting on the P5/P6 dots limits the SPAM fidelity which results in a meaningful reduction in measurement visibility. 
As shown in \reffig{fig:app-2qrb-m1h-m2h}, the decay rates measured on either side are consistent, and we observe similar gate performance from measurement on either side of the device.

Additionally, we have explored the RB sequences with and without $X$-gate prerotations, reminiscent of the single-qubit blind RB protocol~\cite{andrews2019} and of character benchmarking on the larger space of encoded qubits~\cite{helsen2020}.
Unlike the single-qubit case, where linear combinations of the two measurement results can be used to differentiate error in the qubit computational space from leakage error, the two-qubit case navigates through a larger Hilbert space, which requires a greater number of measurements to distinguish those types of errors in a similar manner.
Though we have not yet formalized such a technique, we still emphasize that the decay rates on the singlet and triplet branches are consistent within the confidence intervals.
We take this as a strong sign that the single exponential decay model for each channel is valid here, lending the possibility that linear combinations of exponential decays will enable more detailed leakage analysis in future efforts.

We also note the different asymptotic singlet probability values at large Clifford numbers for the different RB experiments we present.  
These may be understood by a relatively simple counting argument, which utilizes the fact that the total electron spin-projection, $m$, is partially conserved in our experiments.  
This conservation is stronger at higher-magnetic fields, where the mismatch of electron and nuclear Larmor frequencies suppress electron-nuclear flip-flops and pulse sequences are insufficiently fast to drive the spin-system to compensate.  
Although this conservation law is weaker at low-magnetic field, it is strong enough even in Earth's field that the following counting argument applies. 
For 1QRB, there are three states for three spins at constant value of $m_{123}=\pm 1/2;$ in the $\ket{S_{12},S_{123};m_{123}}$ notation, these are  $\ket{0,1/2;m_{123}},\ket{1,1/2;m_{123}},$ and $\ket{1,3/2;m_{123}}$.  
Magnetic noise will, under application of a large number of Clifford sequences, scramble these and result in a probability of 1/3 of measuring $S_{12}=0$ at sequence end.  
For 2QRB, $m$ is the projection across all six spins, and half of experiments begin with $m=0$, while a quarter begin with $m=1$, and another quarter with $m=-1$, depending on the random gauge of the two initialized qubits.  
There are 6!/(3!3!)=20 ways to obtain $m=0$, and 6!/(4!2!)=15 ways to obtain $m=\pm 1$.  
Six of the 20 $m=0$ states have $S_{12}=0$, while 4 of the 15 $m=1$ have $S_{12}=0$, and so if these are again fully scrambled, the expected asymptote is $(6/20)/4+2(4/15)/4=17/60 \approx 0.283$.  
This asymptote is not well observed on M2 in \reffig{fig:app-2qrb-m1h-m2h} and is another indication of reduced SPAM performance on this side of the device.

\section{Physical noise sources}
\label{app:sources}

We now discuss our techniques for characterizing two well-understood sources of noise and decoherence in our device: magnetic noise and charge noise.
This constitutes a more detailed discussion of \reffig{fig:nosc_t2star_brb} than the main text could contain.

Magnetic gradient fluctuations, caused by the noisy magnetization of \Si\ and \Ge\ nuclear spins coupled to the electron spins via the contact hyperfine interaction, provide the most significant contribution to our gate error~\cite{kerckhoff2021}.
While uniform magnetic fields are irrelevant for the DFS encoding, magnetic gradients drive relative precession of electron spin pairs, causing decoherence and leakage out of the encoded subspace.
We characterize this decoherence rate by measuring the decay of a singlet prepared between two electrons and left in the idle configuration for a varying amount of time.
The ensemble-averaged measurement projection decays to a value associated with the predicted mixture of encoded and non-encoded states: 1/2 probability to the initial singlet $\ket{0,1/2;m_{123}}$, 1/6 to the encoded triplet $\ket{1,1/2;m_{123}}$, and 1/3 to the leaked state $\ket{1,3/2;m_{123}}$~\cite{ladd2012}.  
Spin singlet pairs are initialized on either side of the device then shuttled to and from the desired location via consecutive $\pi$ pulses.
Due to the relatively poor SPAM fidelity on the M2 side of the device, spin pairs measured on the M2 side show deviations in the decay asymptote from the predicted value of 1/2.
The gaussian envelope decays with a characteristic timescale \ttwostar, which we measured to be about 3.5~$\mu$s for six different spin pairs, as shown in \reffig{fig:nosc_t2star_brb}(a).   

$T_2^*$ is the most important metric for our error budget, as RB or IRB error scales as $[(\tidle+\tpulse)/T_2^*]^2$ unless decoupling mitigations are applied~\cite{west2012, hickman2013, sun2022}.  
The $T_2^*$ values we observe are in good agreement with theoretical predictions for the known concentrations of \Si\ and \Ge\ isotopes and the silicon well width. 
The reduction of this physical noise source would be accomplished either by employing magnetic decoupling pulse sequences~\cite{west2012, hickman2013, sun2022}, growing samples with reduced \Si\ and \Ge\ content, or by reducing $\tidle$ and $\tpulse$.  
Unfortunately, further reduction to $\tidle$ or $\tpulse$ relative to the values we use here causes additional errors due to pulse distortion effects in our signal chain~\cite{andrews2019}.

Another contribution to our gate error is charge noise, here manifest as exchange noise induced by fluctuations in the lateral trapping and tunneling potentials.
This noise is due either to noise in the signal chain or fluctuating defects in the gate stack and has a strong $1/f$ spectrum across many decades~\cite{connors2021}.
This type of noise induces error only during active exchange pulsing, since nearest-neighbor tunnel coupling is suppressed when the associated spins are idling.  
This idle-error-suppression results from the exponential scaling of the spin-spin exchange interaction with applied barrier potentials in general, and in particular from the large on-off ratios that are available with our tightly-confining SLEDGE design~\cite{haha2021}.
As shown in \reffig{fig:nosc_t2star_brb}(b), we quantify this charge noise contribution to our error budget with an exchange oscillation $Q$-factor at a given $J$ (typically $J/h\sim$100~MHz). 
The product of $J/h$ with the 1/$e$ duration of the gaussian decay envelope of exchange oscillations gives us a number of oscillations \nosc.
$\nosc$ is a superior metric to the actual decay time, because the impact of charge noise on $J$ scales as $|dJ/dV|$ for gate voltages $V$.  
If $J$ were exactly exponential with voltage, \nosc would be independent of voltage, but it is not perfectly so as exchange is subexponential with X-gate voltage~\cite{reed2016}. 
Similar to our magnetic noise heuristic, the estimated theoretical gate error from charge noise scales quadratically with this decay envelope, $1 / \nosc^2$.

We repeatedly measure \nosc\ as a function of $J$ along the symmetric axis, finding that \nosc\  reaches $\sim$50 at 100~MHz, the frequency of operation due to experimental convenience.
Remarkably, and contrary to pure exponential activation, \nosc~rapidly increases from $< 100$ at $J / h = 1$ GHz to $> 1600$ above $J / h \sim 20$ GHz, where we observe a sharp reduction in $|dJ/dV|$ as the potential barrier flattens [\reffig{fig:nosc_t2star_brb}(c)].
The accessibility of this operation point in a SiGe accumulation mode device, where the exchange energy asymptotes to the double-dot orbital splitting and loses sensitivity to control voltages and noise \cite{dial2013}, can be attributed to the large gate action of the SLEDGE design.
To resolve coherent oscillations which occur at exchange frequencies well above the Nyquist frequency of our AWGs (200~MHz), we shift the AWG time basis, which is normally set to 2.5 ns, within the range [2.5,5)~ns. 
A continuous shift of the time basis within this range provides the smooth sampling needed to resolve coherent oscillations without aliasing, and thus extract both the exchange rate, $J/h$, and \nosc. 
While these exchange energies are too high for practical pulsed operation, they may prove valuable for microwave-sensitive EO encodings~\cite{pan2020}.

We perform an initial validation of our error budget with single-qubit randomized benchmarking in \reffig{fig:nosc_t2star_brb}(d).
To properly account for leakage out of our computational space, we use the ``blind'' benchmarking technique described in Ref. \onlinecite{andrews2019}, where sequences are chosen to compile either to the identity or the Pauli-$X$ operation.
We find an average single-qubit Clifford error of $(1.1\pm 0.1)\expn{-3} $ with a leakage error of $(3 \pm 1)\expn{-4}$.
This is in approximate agreement with our simulated prediction of error from magnetic and charge noise alone, $5.0\expn{-4}$, but which does not include the effects of pulse distortion or other physical contributions such as excited state leakage or spin-orbit interactions.

\begin{figure}[htp]
    \includegraphics[width=\columnwidth]{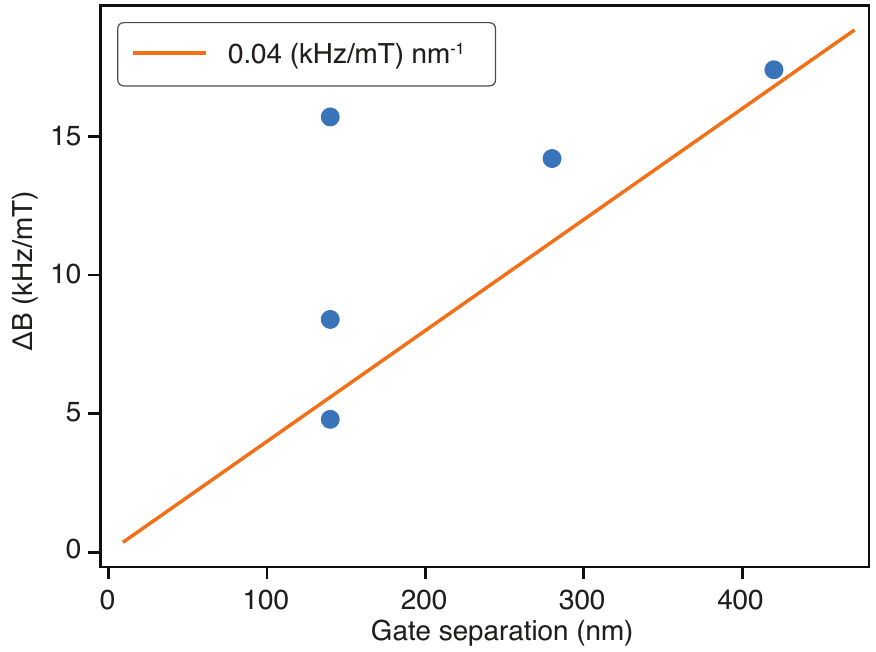}
    \caption{Paramagnetic gradients as a function of gate-to-gate separation in the lateral plane.
    }
    \label{fig:app-paramag}
\end{figure}

As a final note, previous accumulation-mode device designs exhibited significant paramagnetic gradients in response to external applied magnetic fields. 
We attributed such gradients to fringe fields induced by Meissner screening in the aluminum gate stack which, due to the gate stackup complexity, varied in strength and direction over the dot array.
The magnitude of paramagnetic gradients observed here is notably smaller, and consistent with expectations from spin-orbit coupling (see \reffig{fig:app-paramag})  \cite{kerckhoff2021}.
Reducing gradients is a requisite for high fidelity operation in the B $\approx$ 1-10~mT regime, where magnetic gradients would otherwise drive always-on spin-spin interactions detrimental to this control scheme.

\end{document}